\documentstyle[12pt]{article}
\newcommand{\bb}{\begin{equation}}
\newcommand{\ee}{\end{equation}}
\begin{document}
\rightline{hep-th/9506158}
\bigskip
\centerline{\bf QUANTIZATION OF GAUGE THEORIES WITH
ANOMALIES\footnote{Invited Talk at DAE Symposium, Santiniketan, Jan 1, 1995}}
\bigskip\bigskip
\centerline{\it P. Mitra\footnote{e-mail mitra@saha.ernet.in}}
\medskip
\centerline{Saha Institute of Nuclear Physics}
\centerline{Block AF, Bidhannagar}
\centerline{Calcutta 700 064, INDIA}
\bigskip\bigskip
\centerline{\bf Abstract}
\bigskip
In this talk, we briefly review the basic concepts of
anomalous gauge theories. It has been
known for some time how theories with local anomalies can be handled.
Recently it has been pointed out that
global  anomalies,  which  obstruct  the  quantization of certain
gauge theories in the temporal gauge, get bypassed  in  canonical
quantization.
\bigskip\bigskip

Anomalies   of  two  different  types  can  be  involved  in  the
quantization of  gauge  theories.  The  existence  of  divergence
anomalies  has  been  known for a long time \cite{1}: certain classical
theories have symmetry currents which cease to be conserved after
quantization. In case the current is associated with  a  symmetry
which  is  gauged,  there  appears  to  be  a problem in the
quantization of the theory  because the equations of motion of
the  gauge  fields  require  the   current   to   be   conserved.
Fortunately,    these    apparently    contradictory     features
-- nonconservation due to the anomaly and conservation due to the
gauging --   can  be ironed out because the anomaly itself
can  be  made  to  vanish  by going to a submanifold of the classical
phase space before quantization. Of
course, there is a difference  from  theories  with  nonanomalous
gauge  currents. In those theories, there is gauge freedom, which
means that the theories can be described in any  of  an  infinite
variety  of  gauges.  This  is  not  possible in a straightforward
manner in anomalous gauge
theories, where the gauge is, as it were, fixed by the anomaly. However,
we shall see below how an enlargement of the field content can lead
to the usual kind of gauge freedom.

A  second  kind  of  anomaly - the so-called
global, as opposed to the more common local, kind - was discovered in the early
eighties \cite{4}. Here the gauge current is conserved, but  the  group
of   time-independent   gauge   transformations   is  not  simply
connected. This has serious consequences for Dirac quantization  in
the   so-called temporal  gauge.  One  obtains  a representation  of the Lie
algebra of the group of time-independent gauge transformations in
the Hilbert space of states, but this provides  in  general  only
projective (multiple-valued) representations of the group itself.
When  the  fermion content is such that the
representation is not a true one, there is no state in
the Hilbert space which is invariant under the group, so that the
subspace of states obeying Gauss's law is trivial. Fortunately, this problem
can be avoided by fully fixing the gauge. The difference between
theories with global anomalies and anomaly-free theories is
very slight, as we shall see.

We shall review theories with both kinds of anomalies. If one
follows the canonical procedure of quantization, it is easy to see that
there is no conceptual difficulty in quantizing these theories. So we
shall first present this line of argument. But most high energy
theorists nowadays think in terms of functional integrals, so we
shall also explain the difference between anomaly-free theories,
locally anomalous ones and those with global anomalies in the context
of functional integrals.

First we discuss the case of common i.e., local anomalies.
The canonical method of quantization requires the determination of
momenta corresponding to the different field variables. As usual, $A_0$ has no
canonical conjugate, so there is a primary constraint $\Pi_0=0$. To preserve
this constraint in time, it is necessary to have a further constraint,
and this is how Gauss's law appears. In anomaly-free theories, no further
constraint arises, and the above two constraints have vanishing Poisson
brackets, {\it i.e.,} are first class. In anomalous theories, two things
can happen. The preservation in time of Gauss's law may require further
constraints, or alternatively the Poisson bracket of $\Pi_0$ and the Gauss
law operator may be nonvanishing. In the former case, the new constraints
turn out to have nonvanishing Poisson brackets with the above two,
so that one always has second class constraints and there is no gauge
invariance; thus, the new constraints are analogous to the
gauge conditions that one has the freedom to choose in ordinary gauge theories,
so one may say that the gauge is determined by the anomalous theory itself.
In the other case, the closure of the set of second class constraints
at the level of Gauss's law indicates the occurrence of additional degrees of
freedom. Both situations have been seen in the two dimensional chiral
Schwinger model, where there is a regularization parameter $a$ \cite{3}.
For $a=1$,
the number of constraints is exactly as in the anomaly-free case, so it is
a case of the gauge being automatically fixed.
For $a>1$, there are additional degrees of freedom in the form of massless
particles. These may be thought of as would-be gauge degrees of freedom which
have become physical because of the loss of gauge invariance.
Whatever happens, the set
of constraints has to be identified and imposed on the phase space,
quantization being carried out thereafter in terms of the reduced degrees
of freedom. A clarification has to be made here about the occurrence of
effects of the anomaly. The quantization that was referred to above was
the quantization of the gauge field. The fermions are understood to have
been quantized earlier, otherwise the anomaly would not arise.

Next  we  consider  the  problem  arising  in  the  case of global
anomalies. Observe that the argument given above (impossibility of
imposing  Gauss's  law)  is  specific  to
Dirac's method of quantization,
where quantization is done prior to the removal of gauge
degrees of freedom, and is to be contrasted with
canonical quantization \cite{7}, where all constraints
and gauge conditions are imposed
at the classical level and quantization is carried out on the nonsingular
theory.
The imposition of Gauss's  law
and  the  gauge  condition reduces the phase space. The dynamical
system that remains can be quantized as usual. As Gauss's law  becomes  an
operator equation in the Hilbert space, this space does {\it not}
carry  any  nontrivial representation  of  either the gauge group or its Lie
algebra, so  that  there  is  no  question  of  any  complication
involving  projective  representations in canonical quantization.
The  enforcement  of Gauss's law in this approach may  seem  to  be  done  by
brute  force  when compared to Dirac quantization, but the point is that
it works \cite{2}.

We  pass on to  the functional  integral
formulation of the theories. The full partition function of a
gauge theory with fermions will be written as
\bb Z=\int {\cal D}A Z[A],\ee
where $Z[A]$ is the  exponential of the negatived effective action, obtained
by functionally integrating the  exponential of the negatived classical
action over the fermion fields.

In an anomaly-free theory, $Z[A]$ is gauge invariant. The presence
of an anomaly makes $Z[A]$ vary with gauge transformations of $A$:
\bb Z[A^g]=e^{i\alpha (A, g^{-1})} Z[A],\ee
where $\alpha$, which is an integral representation
of the anomaly \cite{6}, has to satisfy some consistency conditions:
\begin{eqnarray}\alpha(A, g_2^{-1}g_1^{-1})&=&\alpha(A^{g_1},g_2^{-1}) +
\alpha(A,g_1^{-1}),\nonumber\\
\alpha(A, g^{-1})&=&-\alpha(A^g,g).\label{con}\end{eqnarray}

The case
of a global anomaly involves  a special form of  $\alpha$.
One way of characterizing a theory with a global
anomaly is to say that the full group of time-dependent
gauge transformations is disconnected.  Thus there is a
possibility of distinguishing between
transformations not  connected  to  the  identity and ones
obtainable from the identity by a sequence of infinitesimal
transformations. It is only under the former, {i.e.}, the
large gauge transformations, that
$Z[A]$  does not stay invariant in these theories. To be precise,
the transformation  is given by
\bb      Z[A^g]=e^{i\gamma (g)} Z[A], \ee
where  $\gamma   (g)$   cannot  be  taken  to  vanish {\it except} for  gauge
transformations $g$ connected to the  identity.

In anomaly-free theories, the full partition function factorizes
into the volume of the gauge group and a gauge-fixed partition function:
\begin{eqnarray} Z&=&\int {\cal D}AZ[A]\nonumber\\
&=&\int {\cal D}AZ[A]\int {\cal D}g\delta(f(A^g))\Delta_f(A)\nonumber\\
&=&\int {\cal D}g\int {\cal D}AZ[A^{g^{-1}}]\delta(f(A))\Delta_f(A)\nonumber\\
&=&\int {\cal D}g\int {\cal D}AZ[A]\delta(f(A))\Delta_f(A)\nonumber\\
&=&(\int {\cal D}g)Z_f.\end{eqnarray}
Here standard Faddeev-Popov notation has been used, with $\delta(f)$
implementing a gauge-fixing condition and $\Delta_f$ the corresponding
Faddeev-Popov determinant. In deriving the fourth equality, the
invariance of $Z[A]$ under a gauge transformation has been used.

The above {\it decoupling of the gauge degrees of freedom} does not occur
if a (local) anomaly is present. In this case, one has \cite{5}
\bb Z=\int {\cal D}g\int {\cal D}A
e^{i\alpha(A,g)}Z[A]\delta(f(A))\Delta_f(A),\ee
in which $g$ and $A$ are seen to be coupled because of the anomaly
term $\alpha$. As indicated above, there are two possibilities: first, it may
happen that there is a gauge function $f^*$ such that
$\alpha$ vanishes in the special gauge $f^*=0$ and so the gauge degrees
of freedom decouple in this gauge; alternatively,
if there is no such gauge, the would-be
gauge degrees of freedom become physical. In this latter case one can still
fix a gauge, but only in an enlarged theory where the field $g$ in the
above expressions is also taken as a dynamical field.
By using the consistency condition (\ref{con}) for $\alpha$, the product
$Z[A,g]\equiv e^{i\alpha(A, g)}Z[A]$ can be seen to be invariant under
gauge transformations of $A$ if $g$ is appropriately transformed at the same
time:
\bb Z[A^h,gh]=Z[A,g].\ee
This is Faddeev's idea \cite{6} of making the anomalous theory gauge invariant
by introducing a Wess-Zumino field to cancel the gauge variation.
The new theory with
\bb Z=\int \int{\cal D}g {\cal D}A
Z[A,g]\delta(f(A))\Delta_f(A),\ee
can be sought to be treated by standard methods. It is
not obvious whether unitarity and renormalizability will hold. Wisdom
gained from experience in two dimensions suggests that there will be
some theories, or more precisely some regularizations, for which
unitarity is violated. Other regularizations have to be used. (It has
to be recognized that the regularization enters the picture through the
form of the anomaly.) The issue of renormalization is less well understood,
because two dimensions cannot help us here.

What about theories with global anomalies? The  partition function
seems to factorize:
\bb Z=\int {\cal D}g
e^{-i\gamma(g)}\int {\cal D}AZ[A]\delta(f(A))\Delta_f(A).\ee
But one has to be careful. The phase factors form a representation
of the group, so
\bb \int {\cal D}ge^{-i\gamma(g)}=
\int {\cal D}(gh)e^{-i\gamma(gh)}=
e^{-i\gamma(h)} \int {\cal D}ge^{-i\gamma(g)}, \ee
where a fixed element $h$ of the gauge group has been used. If it
is not connected to the identity, the left and right hand sides seem
to differ by a phase factor, indicating that
$\int {\cal D}ge^{-i\gamma(g)}$ must vanish.
This   implies   that  the  partition  function
$Z$ vanishes. In fact, this was given as  one  of  the
arguments against the definability of such theories \cite{4}.
However,  one is really interested in  the
expectation values of gauge invariant operators:
\bb{\int{\cal D}AZ[A]{\cal O}\over\int{\cal D}AZ[A]}=
{\int {\cal D}ge^{-i\gamma(g)}
\int{\cal D}AZ[A]\delta(f(A))\Delta_f(A){\cal O}
\over\int {\cal D}ge^{-i\gamma(g)}
\int {\cal D}AZ[A]\delta(f(A))\Delta_f(A)}.\ee
The right hand side is of the form $0/0$ because of the presence of
the factor $ \int {\cal D}ge^{-i\gamma(g)}$ in the numerator and the
denominator. Although it is formally meaningless, one can hope to
interpret this ratio in a sensible way by removing this common vanishing
factor. One thus expects
\bb <{\cal O}>=
{\int{\cal D}AZ[A]\delta(f(A))\Delta_f(A){\cal O}
\over \int {\cal D}AZ[A]\delta(f(A))\Delta_f(A)}.\label{O}\ee

Now (\ref{O}) is precisely what one gets in the {\it canonical} approach to
quantization. We have  considered above the {\it Lagrangian}
functional integral: the singular nature of the
Lagrangian has been ignored and all degrees of freedom, physical or
unphysical,  integrated over. In the canonical approach, on the other
hand, the gauge degrees of freedom are removed
by fixing the gauge at the classical level \cite{7}
and only the physical part of the theory quantized. The functional
integration is then over only the physical fields. There are both
ordinary fields and conjugate momenta, but the latter are easily
integrated over, resulting in functional integrals leading to
(\ref{O}).
This is achieved {\it without} making use of the full partition
function
which was used in the Lagrangian approach and caused all the
problem in this case by happening to vanish.

This simple resolution of the problem
does  not  mean  that  there  is no trace whatsoever of the global
anomaly. An interesting consequence of  the  disconnectedness  of
the  gauge  group  is  that  gauge-fixing  functions  $f$  can be
classified. Two functions $f$ and $f'$ belong to the  same  class
if  one  can  find  a  gauge  transformation  connected to the
identity to go from a gauge field configuration satisfying one
gauge condition to one satisfying the other.
In  this  situation,  $Z_f$  and  $Z_{f'}$ are equal. In general,
however, the  transformation that is needed  will  not  be
connected to the identity. To see what happens in this situation,
we can go through the argument which is used, in anomaly-free
theories, to show that the gauge-fixed partition function is
the same for different gauge functions. Thus,
\begin{eqnarray} Z_f &=&
\int {\cal D}AZ[A]\delta(f(A))\Delta_f(A)\nonumber\\
&=& \int {\cal D}AZ[A]\delta(f(A))\Delta_f(A)
\int {\cal D}g\delta(f'(A^g))\Delta_{f'}(A)\nonumber\\
&=&\int{\cal D}g\int {\cal D}AZ[A]\delta(f(A))\Delta_f(A)
\delta(f'(A^g))\Delta_{f'}(A)\nonumber\\
&=&\int{\cal D}g\int {\cal D}AZ[A^{g^{-1}}]\delta(f(A^{g^{-1}}))\Delta_f(A)
\delta(f'(A))\Delta_{f'}(A)\nonumber\\
&=& \int {\cal D}AZ[A]\delta(f'(A))\Delta_{f'}(A)
\int {\cal D}ge^{-i\gamma(g)}\delta(f(A^{g^{-1}}))\Delta_f(A)
\end{eqnarray}
Were it not for the phase factor $e^{-i\gamma(g)}$, the last integral
would be the identity and the right hand side would reduce to the
gauge-fixed partition function for the gauge function $f'$. The two
integrals appear to be coupled here. But that is not really the case.
Although different gauge field configurations have to be integrated
over, only those are relevant for which both $f'(A)$ and $f(A^{g^{-1}})$
vanish, and the second condition picks out one $g$ for each $A$
satisfying the first condition. As $A$ changes
continuously -- the spacetime is taken to be compactified
-- $g$ varies in  a fixed homotopy class, so that $\gamma(g)$,
which depends only on the class, remains unchanged. Consequently,
the factor can be pulled out and one can write
\bb Z_f=e^{-i\gamma(g_0)}Z_{f'},\ee
where $g_0$ is an element of the relevant homotopy class, which is
determined by the gauge functions $f$ and $f'$. It is through these
factors that theories with  global anomalies differ from anomaly-free
theories. But these factors occur only in the partition functions
and clearly cancel out in the expectation values of gauge invariant
operators, so  that  {\it  Green  functions  of  gauge  invariant
operators are fully gauge independent.}

One   may   wonder   whether   these  theories  are  unitary  and
renormalizable. They  indeed  are.  This  may  be  understood  by
recalling that the gauge current is conserved in  these  theories
-- if  it  is  not,  one  is dealing with a theory with the first
(local) kind of anomaly instead of the second (global) kind!  The
invariance under infinitesimal gauge transformations which exists
in  these  theories  implies  a  BRS  invariance \cite{8}  within  each
gauge-fixed  version.  This,  together  with  the  global   gauge
invariance   of  Green  functions  of  formally  gauge  invariant
objects, may be  used  as  usual  to  demonstrate  unitarity  and
renormalizability \cite{9}.  Indeed, all of standard perturbative gauge
theory  relies  only  on  invariance  under  infinitesimal  gauge
transformations,   {\it   i.e.},   under   the   group  of  gauge
transformations connected to the  identity.  The  theories  under
discussion  do  possess  this  invariance: $\gamma (g)$  is
zero for all $g$ connected to the identity.

There is one mathematical question which has to be addressed: the
possibility of choosing a  gauge  condition  in  these  theories.
There  is,  in  fact,  a  problem,  but  it  applies to all gauge
theories irrespective of whether they are afflicted by  anomalies
of any kind. There is a general theorem \cite{10} asserting that gauges
{\it  cannot} be chosen in a smooth way. But it is also known \cite{10}
that for the construction of functional integrals,
it is sufficient to have  piecewise smooth gauges.
It bears repetition that this has  to  be
done even for theories {\it without} disconnected gauge groups.

To sum up,  the  problems  with gauge theories suffering from
anomalies can be avoided by canonical quantization, where the
singular nature of the gauge field Lagrangian is recognized and all
the constraints  properly imposed. Even the Lagrangian
functional integral approach can be used for both
local and global anomalies if factors of $0/0$ are interpreted in
the most natural way.

\end{document}